# Temperature Sensors based on Semiconducting Oxides: An Overview

B.C. Yadav[1,2]*, Richa Srivastava[1], Satyendra Singh[1], Anurodh Kumar[1] and A.K. Yadav[1]

*Nanomaterials and Sensors Research Laboratory*
Department of Physics, University of Lucknow, Lucknow-226007, U.P., India
[2]*Department of Applied Physics, School for Physical Sciences,*
*Babasaheb Bhimrao Ambedkar University, Lucknow-226025, U.P., India.*

*Email address: balchandra_yadav@rediffmail.com, Mobile: +919450094590

## Abstract

Earlier studies show that organic and inorganic semiconducting materials are the most promising materials for use in temperature sensors. In this brief review, attention will be focused on temperature sensors and its applications in various fields. In addition, we have investigated the temperature sensing characteristics of nanostructured ZnO and ZnO-CuO nanocomposite. For this purpose, ZnO and ZnO-CuO nanocomposite were synthesized via chemical precipitation method. Scanning electron microscopy and X-ray diffraction of sensing materials have also performed. The average crystallite size was 45 and 68 nm for ZnO and ZnO-CuO respectively. The pelletization of the synthesized powder was done using hydraulic pressing machine (MB Instrument, Delhi) under a pressure of 616 MPa at room ambient. This pellet was put within the Ag-Pellet-Ag electrode configuration for temperature sensing. Temperature sensitivities of above semiconducting oxides were calculated. Electrical properties of the materials establish the semiconducting nature of these sensing pellets. In addition, the activation energies of ZnO and ZnO-CuO nanocomposite were estimated.

## 1. Introduction

Temperature is the most often-measured environmental quantity. This might be expected since most physical, electronic, chemical, mechanical, and biological systems are affected by temperature. Some processes work well only within a narrow range of temperatures; certain chemical reactions, biological processes, and even electronic circuits perform best within limited temperature ranges [1-9]. When these processes need to be optimized, control systems that keep temperature within specified limits are often used. Temperature sensors provide inputs to those control systems. Many electronic components can be damaged by exposure to high temperatures, and some can be damaged by exposure to low temperatures. Semiconductor devices and LCDs (liquid crystal displays) are examples of commonly used components that can be damaged by temperature extremes. When temperature limits are exceeded, action must be taken to protect the system. In these systems, temperature sensing helps to enhance the reliability. One example of such a system is a personal computer. The computer's motherboard and hard disk drive generate a great deal of heat. The internal fan helps cool the system, but if the fan fails, or if airflow is blocked, the system components could be permanently damaged. By sensing the temperature inside the computer's case, high-temperature conditions can be detected and actions can be taken to reduce system temperature. Other applications simply require temperature data so that temperature's effect on a process may be accounted for. Examples are battery chargers (batteries charge capacities vary with temperature and cell temperature can help to determine the optimum point at which to terminate fast charging), crystal oscillators (oscillation frequency varies with temperature) and LCDs.

## 1.1. Contact Temperature Sensors

Contact temperature sensors measure their own temperature. One infers the temperature of the object to which the sensor is in contact by assuming or knowing that the two are in thermal equilibrium, that is, there is no heat flow between them. Many potential measurement error sources exist, as you can appreciate, especially from too many unverified assumptions. Temperatures of surfaces are especially tricky to measure by contact means and very difficult if the surface is moving. It is wise to be very careful when using such sensors on new applications. However, all sensors have their own set of complexities.

## 1.2. Thermocouples

Thermocouples are pairs of dissimilar metal wires joined at least at one end, as shown in Figure 1 (A) and (B), which generate a net thermoelectric voltage between the open pair according to the size of the temperature difference between the ends, the relative Seebeck coefficient of the wire pair and the uniformity of the wire-pair. Thermocouples are among the easiest temperature sensors to use and are widely used in science and industry [10-15]. They are based on the Seebeck effect that occurs in electrical conductors, which experience a temperature gradient along their length. They are simple, rugged; need no batteries, measure over very wide temperature ranges.

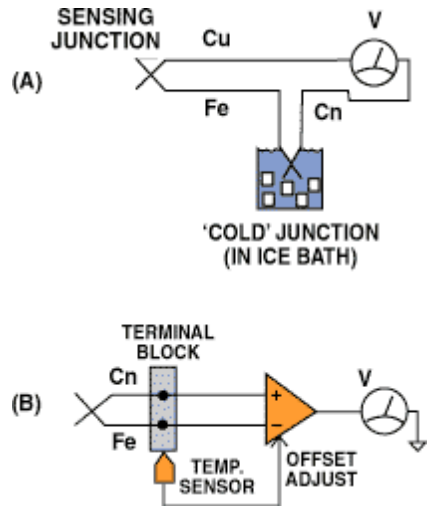

Figure 1. Thermocouple.

Thermocouples measure their own temperature. You must infer the temperature of the object of interest by being certain there is no heat flow between them when you take the measurement. That's easier than it sounds in some case. Thermocouples can make a mistake in reading their own temperature, especially after being used for a while, or if the insulation between the wires loses its resistance due to moisture or thermal conditions, or there are chemical, nuclear radiations or mechanical effects with the immediate surroundings.

Thermocouples are used in many places with many things like indicators and controllers to do something useful, such as control a heating system to heat a product through a temperature-time profile that causes it to soften or cook or set or transform from a stressed condition to an annealed one or any number of physico-chemical changes that produce a desired end result.

### 1.3. Resistance Temperature Detectors (RTDs)

Resistance Temperature Detectors (RTDs) are wire wound and thin film devices that measure temperature because of the physical principle of the positive temperature coefficient of electrical resistance of metals. The hotter they become, the larger or higher the value of their electrical resistance [16-20]. They are most popular, nearly linear over a wide range of temperatures and some small enough to have response times of a fraction of a second. They are among the most precise temperature sensors available with resolution and measurement uncertainties of ± 0.1°C. Usually they are provided encapsulated in probes for temperature sensing and measurement with an external indicator, controller or transmitter, or enclosed inside other devices where they measure temperature as a part of the device's function, such as a temperature controller or precision thermostat.

RTDs can be made economically in copper and nickel, but the latter have restricted ranges because of non-linearities and wire oxidation problems in the case of copper. Platinum is the preferred material for precision measurement because in its pure form the temperature coefficient of resistance is nearly linear; enough so that temperature measurements with precision of ± 0.1°C can be readily achieved. All RTDs used in precise temperature measurements are made of platinum and therefore sometimes called PRTs. The advantages of RTDs include stable output for long period of time, simplicity of recalibration and accurate readings over relatively narrow temperature spans. Their disadvantages, compared to the thermocouples, are: smaller overall temperature range, higher initial cost and less rugged in high vibration environments. They are active devices requiring an electrical current to produce a voltage drop across the sensor that can be then measured by a calibrated read-out device.

### 1.4. Thermistors

Thermistors are special solid temperature sensors that behave like temperature-sensitive electrical resistors [20-32]. Broadly, these are of two types, NTC-negative temperature coefficient thermistors and PTC-positive temperature coefficient thermistors. NTC are used mostly in temperature sensing [33-48], while PTC used mostly in electric current control [49-65].

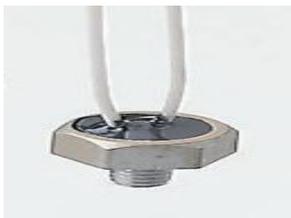
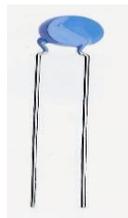
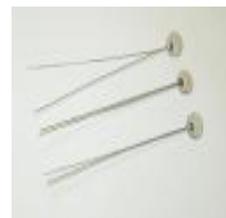

Figure 2(a). A threaded thermistor.
Figure 2(b). A typical disc thermistor.
Figure 2(c). Thermistors.

Different types of thermistors are shown in Figures 2(a), (b) and (c). Temperature sensors are mostly very small bits of special material that exhibit more than just temperature sensitivity. They are highly sensitive and have very reproducible resistance vs. temperature properties. During the last 70 years or so, only ceramic materials was employed for production of NTC thermistors. In 2003, Si and Ge high temperature NTC thermistors were developed with better performance than any ceramic NTC thermistors. Thermistors, since they can be very small, are used inside many other devices as temperature sensing probes for commerce, science and industry. Some of those novel digital medical thermometers that get stuck in one's mouth by a nurse with an electronic display in her other hand are based on thermistor sensors. Thermistors typically work over a relatively small temperature range, compared to other temperature sensors, and can be very accurate and precise within that range.

*1.4.1. Phase Change Thermometers*

Phase change temperature measurement devices or thermometers take many forms and are familiar to lots of people in industry and commerce. Those are found in forms like temperature labels or temperature stickers having a central white or yellowish dot that turns black when the temperature value printed on the label is exceeded.

*1.4.2. Crayon and Paint Thermometers*

There are also temperatures or thermal paints, which change color and temperature crayons that melt and become liquid after the exceed to their specified temperatures. In addition, the thermal paint devices are used only once, but the crayons can be reapplied due to the simple nature of their use.

*1.4.3. Filled System Thermometers*

Filled system thermometers are those that work on pressure or volume change of a gas or changes in vapor pressure of a liquid. The gas or gas and liquid are contained usually in a sealed metal tubing and bulb system. The gas type was used in many industrial applications and for establishing portions of the thermodynamic temperature scale. They can be very simple, non-powered devices with great reliability and repeatability. The vapor pressure types have a bulb, the sensing portion, filled with a volatile liquid, instead of a gas. Since they are more sensitive to temperature changes than a gas type, they can be physically smaller; however, their relative temperature measurement span is quite a bit smaller.

*1.4.4. Bimetallic Thermometers and Thermostats*

Bimetallic thermometers are contact temperature sensors found in several forms if you know where to look, e.g. inside simple home heating system thermostats [66]. They are the coil of metal that has some electrical contacts affixed to it. They are more familiar to many people in industry and commerce as miniaturized pocket dial thermometers that many people use to check the temperature of fat in a deep frier or a vat on a small process line. Perhaps the best part is that the pocket thermometers are sealed and do not require batteries. The major uses are where a quick check of the temperature of an object is desired.

### 1.4.5. *Semiconductors Thermometer Devices*

Commercial temperature sensors have been made from semiconductors for a number of years. Working over a limited temperature range, they are simple, linear, accurate and low cost devices with many uses. Semiconductor thermometers are usually produced in the form of ICs (integrated circuits). There are many types, sizes and models. Most are quite small and their fundamental design results from the fact that semiconductor diodes have voltage-current characteristics that are temperature sensitive. This means that semiconductor triodes or transistors are also temperature sensitive. These devices have temperature measurement ranges that are small compared to thermocouples and RTDs, but, they can be quite accurate and inexpensive and very easy to interface with other electronics for display and control. Semiconductor technology enables devices to be produced efficiently and cheaply and to have properties designed to easily interface with many other types of semiconductor devices, such as amplifiers, power regulators, buffer output amplifiers and microcomputers. The p-n junction is shown in Figure 3(a) and its associated circuitry is shown in Figure 3(b).

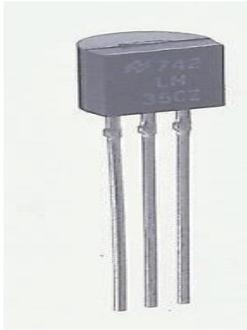
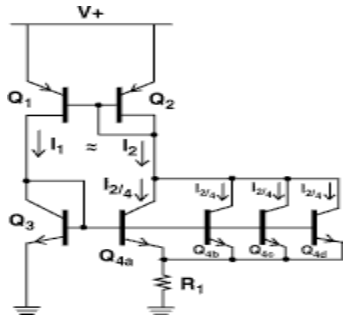

Figure 3 (a). p-n junction sensor.     Figure 3 (b). p-n junction thermometer.

The major uses are where the temperature range is limited to within a minimum temperature of about -25°C to a maximum of about 200°C. Cost, accuracy, simplicity of interfacing with other circuit elements and size are factors in selecting a device to do a job and meet the requirement of both the accuracy and cost budgets.

### 1.5. Other Types of Temperature Sensors

### 1.5.1. *Acoustic and Ultrasonic based Temperature Sensors*

Several groups have exploited the concept of measuring temperature in a gas by measuring the speed of sound in that gas. Another variation on the same idea has been to send ultrasonic pulse down a rod of known expansion and propagation properties. By placing slots in the rod at known and calculable distances from the excitation position, one could immerse the rod in a medium of high temperature and then measure that temperature by measuring the reflection times of the pulses from the notches.

### 1.5.2. *Non-contact Temperature Sensors*

Surface temperature measurement problem can be solved in many cases through the use of non-contact sensors; they are almost ideal for those types of applications and are in use in many industrial plants worldwide in great numbers. In many industrial plants, noncontact sensors are not yet standardized to the extent that thermocouples and RTDs are. In spite of this, there are numerous showcase uses of them and they pay more than their way in process plants such as steel, glass, ceramics, forging, heat treating, plastics, baby diapers and semiconductor operations. More recently, the medical world has adopted the IR ear thermometer that is a single waveband radiation thermometer. They all are based on Planck's law of the thermal emission of radiation.

### 1.5.3. *Radiation Thermometers*

Radiation Thermometers are non-contact temperature sensors that measure temperature from the amount of thermal electromagnetic radiation received from a spot on the object of measurement [67]. This group of sensors includes both spot or point measuring devices in addition to line measuring radiation thermometers, which produce 1-D and, with known relative motion, can produce 2-D temperature distributions, and thermal imaging, or area measuring, thermometers which measure over an area from which the resulting image can be displayed as a 2-D temperature map of the region viewed. These are significant devices in all their manifestations because they enable improvements in processes, maintenance, health and safety that save both lives and money. They are used widely in many manufacturing process like metals, glass, cement, ceramics, semiconductors, plastics, paper, textiles, coatings, and more. They enable automation and feedback control that boost productivity while improving yield and product quality.

### 1.5.4. *Optical Pyrometers*

The optical pyrometer is highly developed and well accepted noncontact temperature measurement device with a long and varied past from its origins more than 100 years ago. Optical pyrometers work on the basic principle of using the human eye to match the brightness of the hot object to the brightness of a calibrated lamp filament inside the instrument. The optical system contains filters that restrict the wavelength-sensitivity of the devices to a narrow wavelength band around 6500 to 6600 Å microns. Other filters reduce the intensity so that one instrument can have a relatively wide temperature range capability. Needless to say, by restricting the wavelength response of the device to the red region of the visible, it can only be used to measure objects that are hot enough to be incandescent, or glowing. This limits the lower end of the temperature measurement range of these devices to about 700 °C. Modern radiation thermometers provide the capability to measure within and below the range of the optical pyrometer with equal or better measurement precision plus faster time response, precise emissivity correction capability, better calibration stability, enhanced ruggedness and relatively modest cost.

### 1.5.5. *Thermal Imagers*

Thermal imagers are a special sub-class of thermal imaging devices; they measure radiation temperature distributions as well as shown a false color thermal image. They are basically single waveband radiation thermometers that measure a two dimensional space instead of just radiation from a single spot.

### 1.5.6. *Fiber Optic Thermometers*

There are a wide number of devices that utilize fiber optics to aid in measuring temperature [68-74]. GaAs semiconductor temperature sensor is shown in Figure 4.

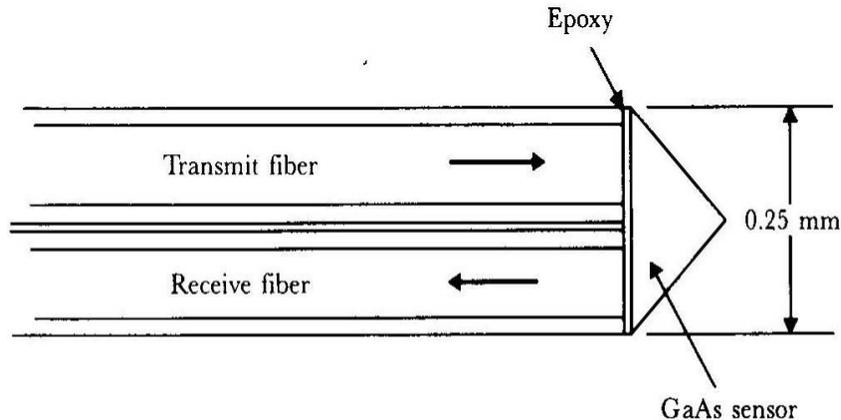

Figure 4. GaAs semiconductor temperature probe.

## 2. Temperature Sensors: Uses

### 2.1. *Cryogenics*

It is a very important area in basic science, engineering, food, metallurgy, and manufacturing. It deals with cold and colder, essentially everything below about -150°C or 123 K. Most people associate this region with liquid nitrogen and liquid oxygen; some even with liquid hydrogen and liquid helium, for these are some of the gases used as cooling agents or propellants in many space vehicle rocket engines. Much of the work reported in the popular press does indeed revolve around work in these areas, e.g. 77 K to 4 K.

**Some uses of Cryogenics**
- Liquified gases, generation, storage and handling
- Physics research
- Hall Effect/Magnet Studies
- Materials and metallurgical research
- Ceramics research
- Carbon research
- Fuels research
- Nuclear Magnetic Resonance (NMR)
- Low temperature research
- Semiconductor laser development
- Superconductor research & development

- Tritium liquifaction
- Nuclear physics detectors

## 2.2. *Food Applications*

The temperature of food plays a big role in assuring that certain products are well enough cooked to kill harmful organisms like bacteria. Similarly, many foods, including cooked food, become breeding grounds for other harmful organisms if unrefrigerated too long or even if left in a refrigerated environment for too long a time [75]. Certainly, in the cooking area it is quite straightforward to monitor the internal temperature of meats and other foods to assure that the proper minimum temperature has been attained before it is considered safe.

## 2.3. *HVAC/R Applications*

Heating Ventilating, Air Conditioning and Refrigeration (HVAC/R), is big business with large companies providing equipment, design services and installation. Small and large companies provide many support services. Many companies and individuals of many organizations help oil the wheels of commerce in this area with a variety of supporting products and services that range from specialty sensors to software to consulting and training. It is an activity that consumes a very large number of temperature sensors and humidity sensors. It is an area of activity that has many highly trained people, like engineers and scientists that develop and engineer equipment and projects.

## 2.4. *Steel and Metals Applications*
- Taconite temperatures in pelletizing operations
- Sinter temperature
- Coke oven temperature measurements and transfer belts protection
- Thermocouples in Blast Furnace Environments
- Stove Domes and Bustle Pipe Temperatures
- Temperature Measurement of Liquid Iron, Liquid Steel and other molten metals
- Slag detection in steel pouring streams and detection of Iron in slag streams
- IR Radiation Thermometers for oxidized steel objects in cooler surroundings
- Non-Contact measurement of steel surface temperatures in Reheat Furnaces
- IR Radiation Thermometers used in Continuous Anneal Furnaces
- Non-contact temperature measurement on Coating Lines, e.g. tin, zinc, plastic film
- Measuring steel sheet surface temperatures in the Galvanneal process

## 2.5. *Medical Applications*

Ever since one's first experience with a fever thermometer, almost everyone has known what it means "to have a temperature". The fever thermometer is still in abundant supply and still the second most frequent test used (after a hand on the forehead) to indicate the presence of an infection in the human body by noting an elevation in body temperature. Human and animal body temperatures are so important [76-79] to the well being of warm-blooded animals, that the nominal body temperature indicated by a fever thermometer or similar device is used as one of the vital signs routinely monitored as an indicator of a state of a person or animal's health.

## 3. Applications of nanostructured semiconducting oxides as Temperature sensor

Earlier studies show that organic and inorganic semiconducting materials are the most promising materials for temperature sensors [80-84]. As the temperature is an important parameter for measuring the different properties of materials, therefore, due to various useful properties of ZnO, it behaves as good temperature sensor [84]. Sucrose shows semiconducting nature which may be utilized for temperature sensor [85]. The copper (I, II) oxides have a special importance in the application of temperature sensors. Previously we have investigated the temperature sensing characteristics of cuprous oxide. The average temperature sensitivity of cuprous oxide was found .74 MΩ/ °C.

In the present work, extensive experimental investigations of ZnO and ZnO-CuO nanocomposite have been studied as temperature sensor. For this purpose pellets of these sensing materials have been made and exposed to temperature. The variations in resistance of the sensing elements with temperatures were recorded. Initially, the resistance of the sample was decreased drastically with increasing temperature, after that it became almost constant. Activation energies of both sensing materials were calculated.

### 3.1. Experimental:

### 3.1.1. *Synthesis of sensing materials:*

ZnO was prepared by conventional precipitation method by adding sodium hydroxide solution to zinc sulphate solution (1:2.2) using 'sudden mixing method' [86]. CuO was synthesized using Benedict's reagent and Benedict's reagent was prepared by dissolving 173 g of sodium citrate and 90 g of anhydrous sodium carbonate in 500 mL of deionized distilled water. Separately 17.3 g of $CuSO_4.5H_2O$ was dissolved in 150 mL of distilled water. This solution was added slowly with stirring to the above solution, the mixed solution was ready for use. Now a mixture was prepared by adding 50 mL glucose solution (0.2 g/mL) to 150 mL Benedict's reagent and then it was boiled for 5 min. The proportion was set to give maximum yield of cuprous oxide particles and nearly full consumption of Benedict's reagent and reducing glucose so that we could find highly pure $Cu_2O$. After cooling, the boiled mixture, brick red precipitate of cuprous oxide, was obtained, which settled down within few minutes. Now filtration of cooled solution gave precipitate of cuprous oxide particles. This precipitate was washed many times by deionized distilled water. The precipitate was dried slowly and brick red fine powder of cuprous oxide was obtained.

The obtained oxides of copper and zinc were mixed in equal proportion through a solving agent. The resulting solution was stirred for 6 h at 50ºC and then filtered. The obtained precipitate was washed with distilled water, and then dried overnight. Further, it was annealed at 450ºC for 2 h.

Each circular pellet having diameter 10 mm and thickness 4 mm was made by using hydraulic pressing machine (M.B. Instruments, Delhi) under the pressure of 616 MPa at room temperature. Further the pellet was put within Ag-Pellet-Ag electrode configuration as shown in the Figure 5 and this configuration was put inside the electrical furnace for temperature sensing and variations in resistance with temperature were recorded. The used heating rate was 2°C/minute.

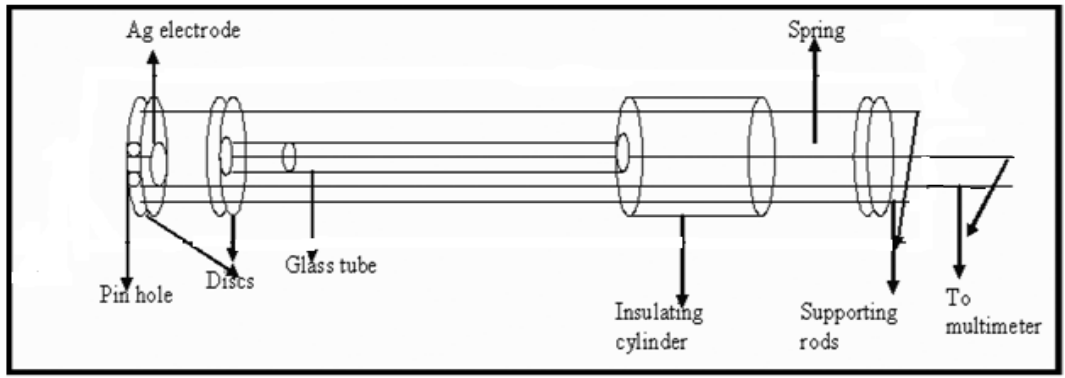

Figure 5. Sample holder (Ag-pellet-Ag electrode configuration).

### 3.1.2. *Characterization of the samples:*

#### 3.1.2.1. *Scanning electron microscopy analysis:*

The surface morphology of the pellet surface was obtained from the Scanning electron microscope (SEM, LEO-0430 Cambridge) and is visualized in Figures 6 (a) and (b), for ZnO and ZnO-CuO nanocomposite respectively. Figure 6 (a) shows that the surface of the ZnO pellet is more porous, which provides larger specific surface area for the adsorption of atmospheric oxygen. While dense and compact structure was seen for ZnO-CuO nanocomposite.

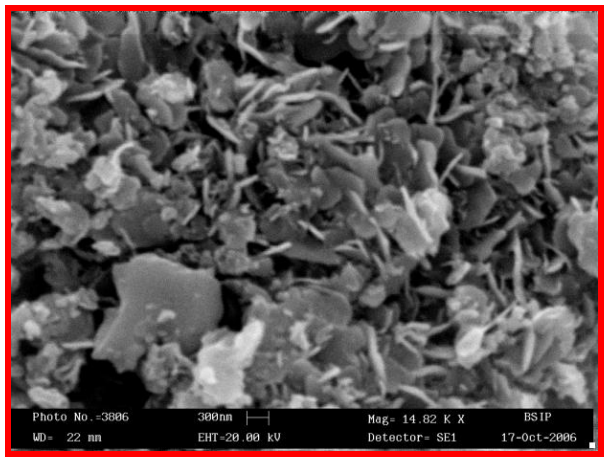 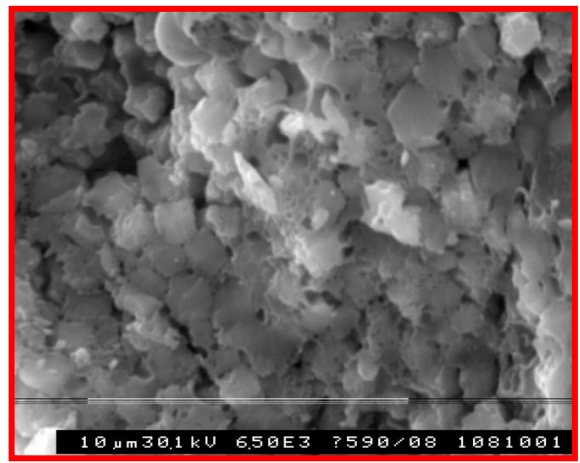

Figure 6(a): SEM image ZnO.　　　　Figure 6(b): SEM image of ZnO-CuO nanocomposite.

The surface morphology can be well tuned by controlling various parameters during synthesis and fabrication of sensing elements such as concentration of solution, pellet thickness, drying and calcination times and temperatures. Generally during the synthesis, calcinations temperature affects the surface morphology of the material. The adsorption phenomenon largely depends on the surface area and it also depends on the size of the particles under investigation. Since sensing surface has dangling bonds, therefore, surface can be chemically very reactive for adsorption of atmospheric oxygen.

### 3.1.2.2. *X-ray diffraction analysis:*

The crystal structure and phase of the powdered material was analyzed using X-ray Diffractometer (X-Pert, PRO PANalytical XRD system, Nether land) with Cu $K_\alpha$ radiations as source having wavelength 1.5418 Å. XRD pattern of the synthesized powder shown in Figures 7(a) and (b), for ZnO and ZnO-CuO nanocomposite respectively. Figure 7(a) reveals the formation of zinc oxide with intense peak centered at 36.5° corresponding to plane (101). The average crystallite size was found 45 nm.

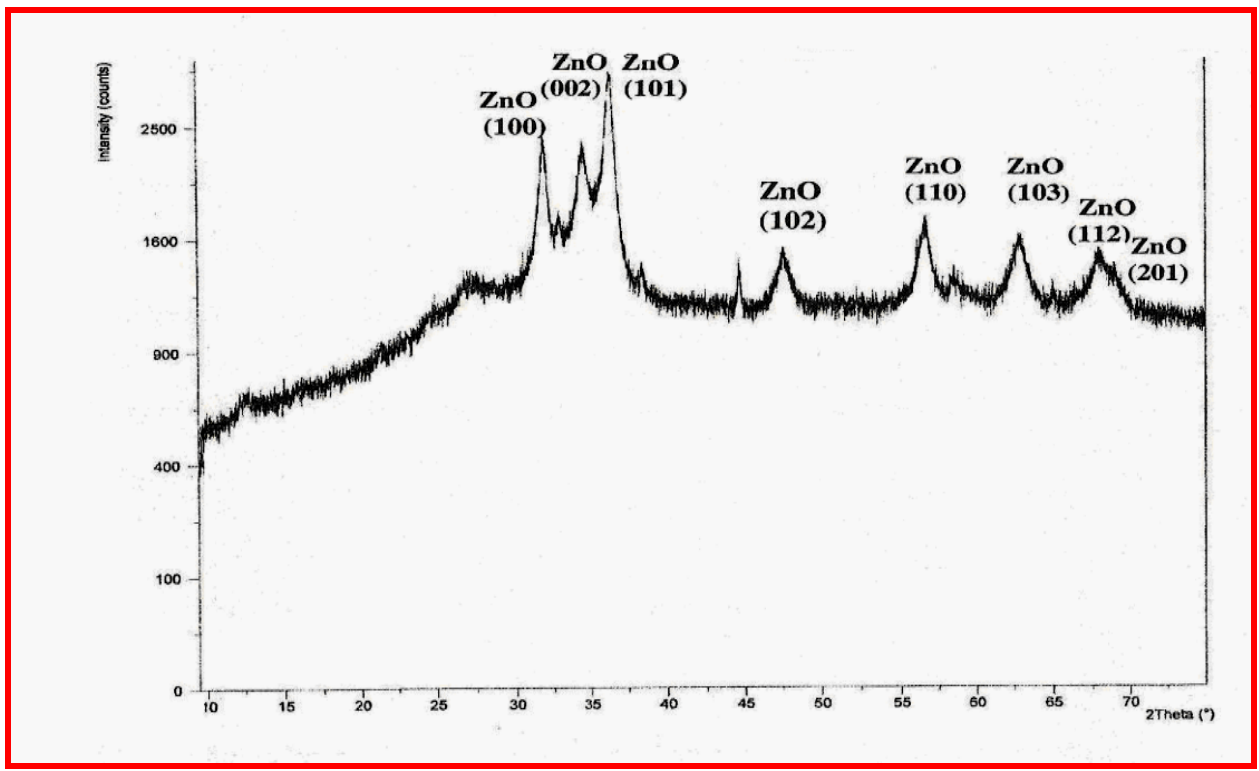

Figure 7(a): XRD pattern of ZnO.

Figure 7 (b) shows the XRD of ZnO-CuO nanocomposite. The peaks of cuprous oxide ($Cu_2O$), cupric oxide (CuO) and zinc oxide are present in the XRD pattern. The highest insensitive peak exists at 36.5° corresponding to the plane (101) and is of ZnO. The average crystallite size was found 68 nm estimated by Debye-Scherrer formula.

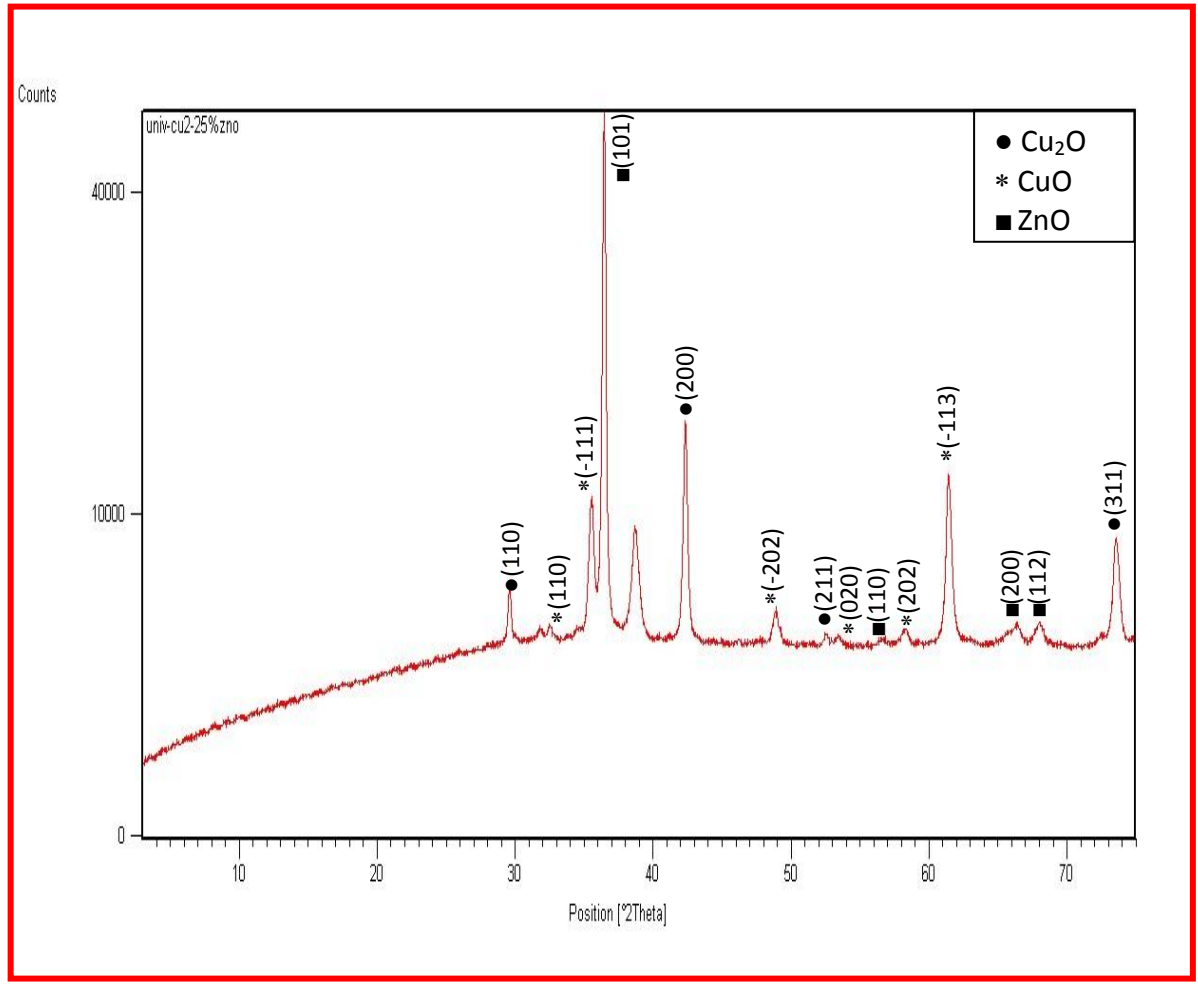

Figure 7(b): XRD pattern of ZnO-CuO nanocomposite.

### 3.1.3. *Resistance-Temperature characteristics (Temperature sensitivity):*

The sensitivity of temperature sensor can be defined as

$$S = \Delta R/\Delta T \quad M\Omega/°C$$

Where $\Delta R$ is change in resistance in $M\Omega$, and $\Delta T$ is change in temperature in °C of sensing materials. We have plotted the variations in resistances with increase in temperature for the sensing elements prepared at room temperature and are shown in Figures 8 and 9. Figure 8 shows the variations in resistance with temperature for sensing element made of ZnO, which indicates the drastic decrease in resistance with increase in temperature for lower range of temperature up to 200°C. Further, as the temperature increases the decrease in resistance becomes lesser and lesser and finally it approaches an almost constant value. The average sensitivity for ZnO was found 1.2 MΩ/°C. The variations in resistance for sensing element made of ZnO-CuO is shown in Figure 9 and it represents that there is similar decrease in resistance and average sensitivity was 0.8 MΩ/°C. Thus, the average sensitivity of ZnO-CuO was found lesser than that of the sensor made of ZnO. Therefore, it was found that the temperature sensor made of ZnO is more sensitive than the sensor made of ZnO-CuO.

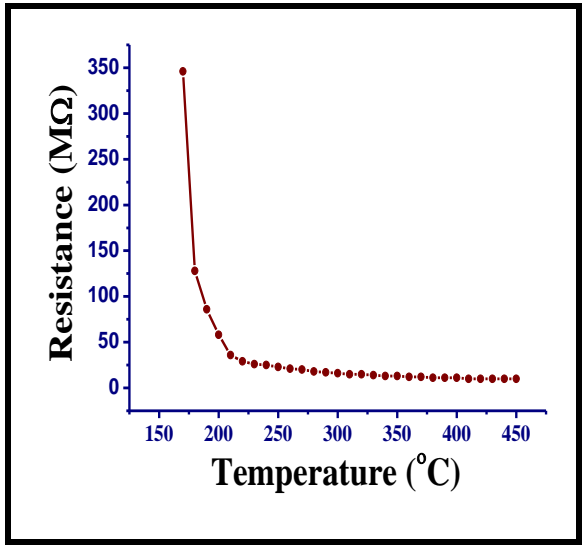
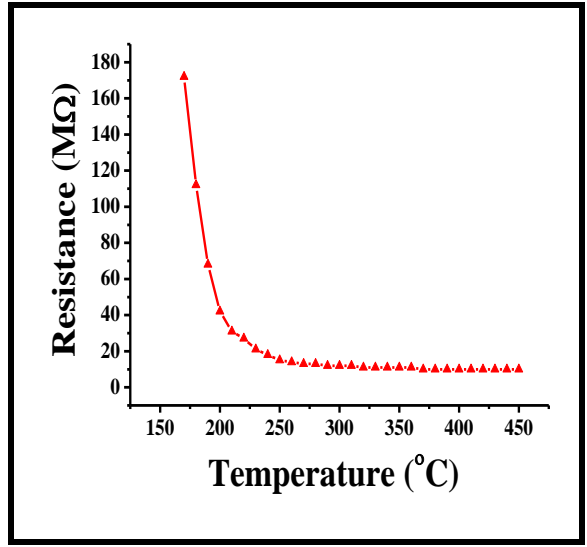

Figure 8. Variations in resistance with temperature for ZnO.

Figure 9. Variations in resistance with temperature for ZnO-CuO nanocomposite.

The physical property that defines a semiconductor is its decrease in its electrical resistance with increasing temperature. Therefore, Figures 8 and 9 show the semiconducting nature of these sensing materials. In these figures the decrease in resistance with the temperature must mainly regarded as due to the thermally activated mobility of the carriers rather than to a thermally activated generation of these.

## 4. Activation energy by thermal resistance method:

Activation energy (ΔE) measures the thermal or other form of energy required to raise electrons from the donor levels to the conduction band or to accept electrons by the acceptor levels $E_a$ from the valence band respectively for n- and p- type materials. Activation energy corresponds to the energy difference ($E_c$ - $E_d$) and ($E_a$ - $E_v$) respectively for n- and p- type semiconductors as shown in Figures 10 and 11. Activation energy by thermal resistance method can be measured from the variation of σ or ρ and conveniently of R with the temperature.

The temperature dependence of conductivity for a semi conducting material can be obtained using following simplified expressions: $\sigma = \sigma_0 e^{(-\Delta E/2KT)}$

Since dimensions remain the same during small temperature variations, therefore, the equation can be simplified for resistance as $R = R_0 e^{(\Delta E/2KT)}$

Where $\sigma_o$, $\rho_o$, or $R_o$ are the pre-exponential factors and is given by

(i) For p-type semiconductor, $\sigma_0 = 2e\mu_h(2\pi m_h^* KT/h^2)^{3/2}$

and activation energy for the excitation of acceptor atoms, $\Delta E = E_a - E_v$

(ii) For n-type semiconductor, $\sigma_0 = 2e\mu_e(2\pi m_e^* KT/h^2)^{3/2}$

and activation energy for the excitation of donor atoms, $\Delta E = E_c - E_d$

(iii) For intrinsic semiconductor

$$\sigma_0 = 2e(\mu_e * \mu_h)(m_e^* m_h^*)^{3/4}(2\pi KT/h^2)^{3/2}$$

and activation energy, $\Delta E = \dfrac{E_c - E_v}{2} = \dfrac{E_g}{2}$

Here $m_e^*$, $m_h^*$ are effective masses and $\mu_e$, $\mu_h$ are mobilities of electrons and holes respectively, $E_c$ and $E_v$ are energy values corresponding to bottom-edge of the conduction band and the top edge of the valence band respectively, $E_g$ is the band gap of the semiconductor, T is the absolute temperature of the material and K is the Boltzmann constant.

The temperature resistance plot in the form of ln ρ and (1/T), known as Arrhenius plot, has a slope of (ΔE/2K) according to equation [87]:

$$\ln \rho = \ln \rho_0 + \Delta E/2KT$$

By measuring the slope of Arrhenius plot of a linear zone, we have calculated the activation energy of ZnO and ZnO-CuO nanocomposite.

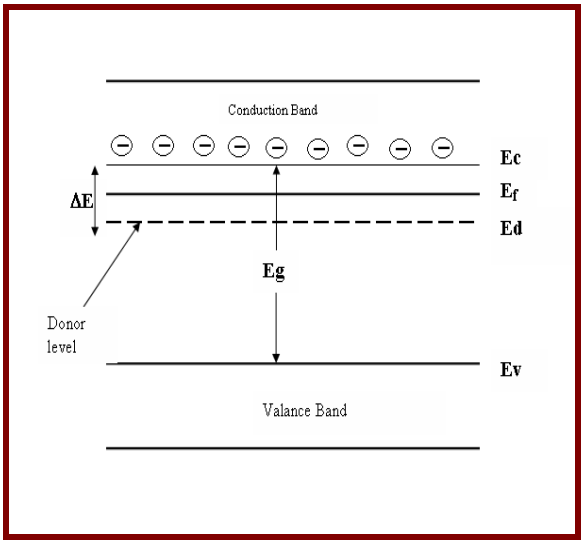
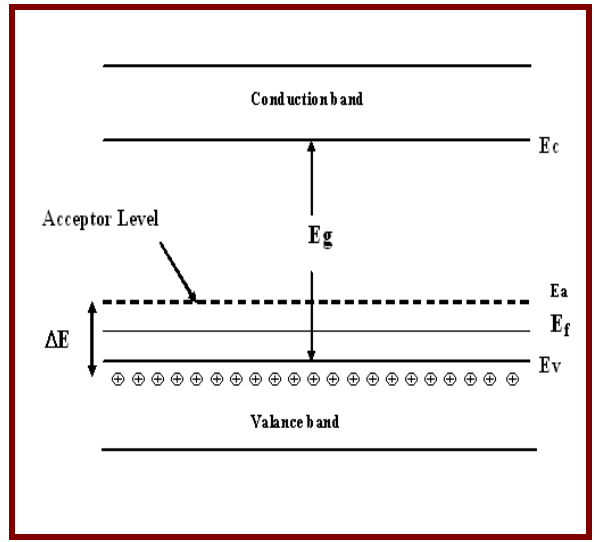

Figure 10: Band diagram for Activation Energy in n-type semiconductor.

Figure 11: Band diagram for Activation Energy in p-type semiconductor.

Figures 12 and 13 show the variation of ln R as a function of inverse temperature for the synthesized ZnO and ZnO-CuO nanocomposite, respectively. The resistance variation of the ZnO and ZnO-CuO nanocomposite can be ascribed to typical band conduction. It can be noted that a change in temperature will alter the resistance because both the charge of the surface species ($O_2$, $O_2^-$, $O^-$ or $O^{2-}$) as well as their coverage can be altered in this process. Since the conduction process in metal oxide semiconducting materials depend heavily on grain boundaries therefore large and small particle sizes of materials are responsible for deviation from straight line behavior.

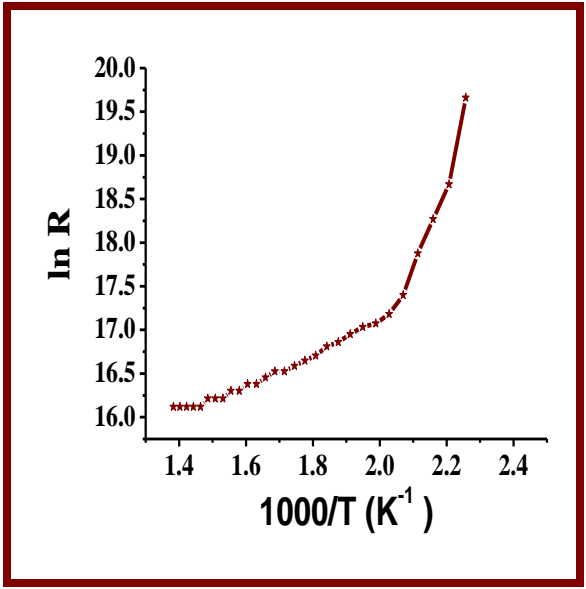
Figure 12: Arrhenius plot for ZnO.

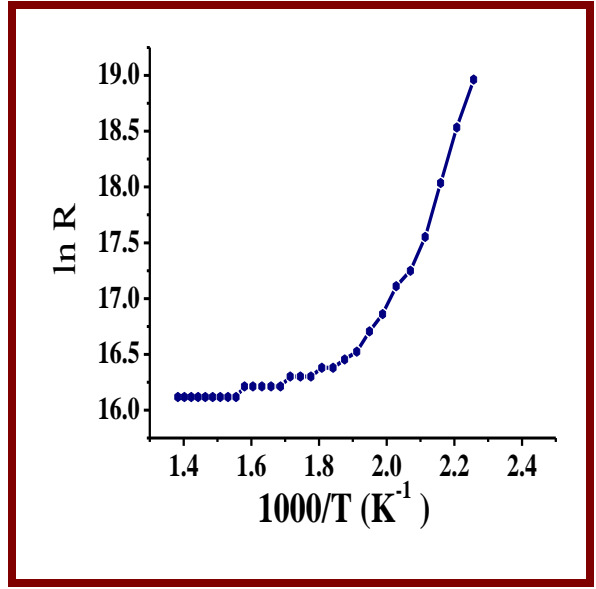
Figure 13: Arrhenius plot for ZnO-CuO.

In the overall conduction process a contribution arising from the participation of ZnO and ZnO-CuO nanocomposite with lower average particle size and another with higher average particle size i.e., the distribution of particle size dominates in thermally activated conduction process.

The energy transition in an investigated temperature interval (170-450ºC), which may be an electron excitation from valence band to an acceptor level, creates a hole in valence band for the conduction. Therefore this transition controls the R-T characteristics. The activation energies determined from the slope of resistance data was found 1.55 and 1.09 eV respectively for ZnO and ZnO-CuO nanocomposite. Thus we conclude that pure ZnO is a better temperature sensor than ZnO-CuO nanocomposite. The values of average sensitivities for ZnO and ZnO-CuO were found 1.2 and 0.8 MΩ/°C, respectively.

## 5. Conclusion:

This review summarizes the types of temperature sensors and their applications in various fields. It describes the significance of temperature as the most often-measured environmental quantity, types of temperature sensors and their applications. Contact temperature sensors measure their own temperature. Thermocouples are among the easiest temperature sensors to use and are widely applied in science and industry. Resistance temperature detectors are wire wound and thin film devices that measure temperature. Thermistors are special solid temperature sensors that behave like temperature-sensitive electrical resistors. Phase change temperature measurement devices or thermometers take many forms and are familiar to lots of people in industry and commerce. Bimetallic thermometers are contact temperature sensors found in several forms e.g. inside simple home heating system. Semiconductor thermometers are usually produced in the form of ICs (integrated circuits). Surface temperature measurement problem can be solved in many cases through the use of non-contact sensors. Radiation thermometers are non-contact temperature sensors that measure temperature from the amount of thermal electromagnetic radiation received from a spot on the object of measurement. The optical pyrometer is a highly developed and well-accepted noncontact temperature measurement device with a long and varied past from its origins more than 100 years ago. Fiber optic thermometers utilize fiber optics to aid in measuring temperature. The use of temperature sensors for particular temperature sensitive range was discussed. In addition the role of semiconducting oxides for temperature sensing was discussed. In particular for ZnO and ZnO-CuO, we have plotted resistance-temperature sensing curves. The temperature sensitivities for ZnO and ZnO-CuO were found 1.2 and 0.8 MΩ/°C, respectively. The estimated values of activation energies for ZnO and ZnO-CuO were 1.55 and 1.09 eV respectively. Thus, we can say that semiconducting oxides are promising materials for temperature sensing. Currently metal oxide semiconducting sensors find limited commercial use; other types of temperature sensors are still favored for many applications.

7. Acknowledgement:

**Dr. Richa Srivastava** is highly grateful to University Grants Commission, Delhi for Post Doctoral Fellowship (No. F.15-79/11 (SA-II)).